\documentclass[%
reprint,
showpacs,preprintnumbers,
amsmath,amssymb,
aps,
prb,
]{revtex4-1}

\usepackage{graphicx}
\usepackage{dcolumn}
\usepackage{bm}
\usepackage{colortbl}

\begin{document}


\title{Fluid photonic crystal from colloidal quantum dots}

\author{V.\,N.\,Mantsevich$^{1}$}
\email{vmantsev@gmail.com}
\author{S.\,A.\,Tarasenko$^{2}$}

\affiliation{%
$^{1}$Moscow State University, 119991, Moscow, Russia,
$^{2}$Ioffe Institute, 194021, St. Petersburg, Russia }%

\date{\today }
\begin{abstract}
We study optical forces acting upon semiconductor quantum dots and
the force driven motion of the dots in a colloid. In the spectral
range of exciton transitions in quantum dots, when the photon energy
is close to the exciton energy, the polarizability of the dots is
drastically increased. It leads to a resonant increase of both the
gradient and the scattering contributions to the optical force,
which enables the efficient manipulation with the dots. We reveal
that the optical grating of the colloid leads to the formation of a
fluid photonic crystal with spatially periodic circulating fluxes
and density of the dots. Pronounced resonant dielectric response of
semiconductor quantum dots enables a separation of the quantum dots
with different exciton frequencies.
\end{abstract}


 \maketitle

\section{Introduction}

Optical fields interacting with micro- and nanoparticles induce sizeable mechanical forces
acting upon the particles, which enables their trapping and precise manipulation~\cite{Ashkin1970,Ashkin1975}.
Such non-invasive optomechanical approaches have been demonstrated for dielectric\cite{Burns1990,Cizmar2006,Yang2009}, metallic\cite{Svoboda1994,Hansen2005,Huang2015,Petrov2016}, semiconductor\cite{Pan2007,Jauffred2008,Jauffred2010} particles, as well as biological cells~\cite{Ashkin1987,Wang2011,Zhong2013}. Using photonic interference schemes, one can create the arrays of optical traps and form the lattices of dielectric particles~\cite{Grigorenko2008} or biomolecules~\cite{Soltani2014}. Among other achievements of optomechanics are measurements of interaction forces between molecules~\cite{Meiners2000} and optical forces with femtonewton resolution~\cite{Zensen2016}.

Highly interesting objects for optical trapping and manipulation are
colloidal nanocrystals (or quantum dots,
QDs),~\cite{Pan2007,Jauffred2008,Jauffred2010} since they are
important for sensor and biological applications, particularly as
superior fluorescent
labels~\cite{Dubertret2002,Dahan2003,Ebenstein2009,Yum2009}. The
optical force acting upon a QD is determined by the dot
polarizability~\cite{Chaumet2000} which has a resonant behavior at
the frequency of exciton transitions in the QD, when the radiation
excites electron-hole pairs (excitons) in the
dot~\cite{Wang2006,Ivchenko_book}. Far from the exciton resonance,
the polarizability is given by the background dielectric contrast
between the QD material and  the environment and commonly described
by the Clausius-Mossotti relation. Optical trapping of colloidal
CdTe- and CdSe-core QDs in such conditions has been demonstrated in
Refs.~\onlinecite{Pan2007,Jauffred2008,Jauffred2010}. Close to the
exciton resonance, the QD polarizability and, correspondingly, the
optical force are drastically enhanced. Currently, there is a lack
of theoretical studies of this effect as well as the optical force
driven dynamics of quantum dots in the colloid.

Here, we study the optical force acting upon a QD and the force
driven fluxes of QDs in a colloid induced by optical grating. We
consider that the dielectric response of a QD is caused by exciton
transitions and has a resonant behavior. The force can be split in
two contributions: the gradient force and the scattering/absorption
force determined by the real and imaginary parts of the
susceptibility, respectively~\cite{Chaumet2000,Rohrbach2001}. Close
to the excitonic resonance, both contributions increase and are
comparable to each other. In the presence of both contributions, the
total force is non-conservative and cannot be described by a
potential. Therefore, the optical grating of a colloid leads to the
formation of spatially periodic density of the dots as well as the
emergence of spatially periodic circulating fluxes of the dots
(Brownian vertexes~\cite{Sun2009}). The emerging modulation of the
QD density and fluxes results, in turn, in the periodic modulation
of the optical properties of the colloid, such as the refractive
index. As a result, the system demonstrates the properties of a
photonic crystal. By considering the interplay of optical forces and
the processes of viscous friction and diffusion of dots, we study
the occurrence of such a fluid photonic crystal from colloidal dots.
We calculate the steady-state distributions of the dot density and
fluxes as well as the time scales of the photon crystal formation.

\section{Theory}
\subsection{Optical force acting upon a quantum dot}

\begin{figure}[t]
\includegraphics[width=60mm]{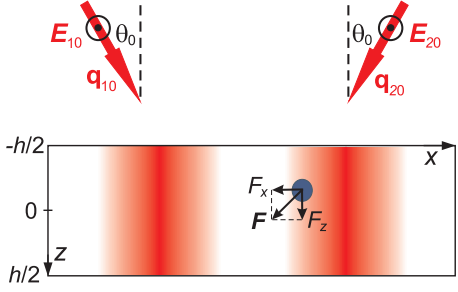}%
\caption{(Color online) Optical grating in a colloidal solution of
quantum dots induced by two coherent laser beams leads to the
emergence of a position-dependent optical force acting upon quantum
dots. Color intensity illustrates the distribution of radiation
intensity in the cell with the colloid, $h$ is the depth of the
cell.} \label{figure1}
\end{figure}

We consider a cell with a colloidal solution of semiconductor QDs
excited by two coherent laser beams of the $s$-polarization and the same frequency $\omega$, as shown in Fig.\,\ref{figure1}.
The laser fields interfere and produce an optical grating. The total electric field of the radiation
in the solution is given by $\bm{{\cal E}}(\textbf{r},t) = {\rm Re} [ \textbf{E}(\textbf{r}) {\rm e}^{-i \omega t} ]$, where
\begin{eqnarray}\label{field}
\textbf{E}(\textbf{r}) = (E_{1}  {\rm
e}^{i\textbf{q}_{1}\cdot\textbf{r}} + E_{2} {\rm
e}^{i\textbf{q}_{2}\cdot\textbf{r}}) \hat{\textbf{y}} \,,
\end{eqnarray}
$E_1$ and $E_2$ are the (real) amplitudes of the laser electric
fields in the solution, which are determined by the amplitudes of
the incident fields, $E_{10}$ and $E_{20}$, respectively, and the
Fresnel transmission coefficient, $\textbf{q}_{1} = q \, (\sin
\theta, 0, \cos \theta)$ and $\textbf{q}_{2} = q \, (-\sin \theta,
0, \cos \theta)$ are the wave vectors, $q = \omega n_{\omega} /c$,
$\theta$ is the angel of refraction, which is related to the angle
of incidence $\theta_0$ by Snell's law
$\sin\theta_0=n_\omega\cdot\sin\vartheta$, $n_{\omega}$ is the
refractive index of the solution, $c$ is the speed of light, and
$\hat{\textbf{y}}$ is the unit vector along the $y$ axis. The
intensity of the radiation $I$ for the geometry we consider depends
on the $x$ axis only and is given by
\begin{equation}\label{intensity}
I = \frac{c n_{\omega}}{8\pi} |\textbf{E}|^{2} = \frac{c
n_{\omega}}{8\pi} [E_1^2 + E_2^2 + 2 E_1 E_2 \cos(2 x q \sin
\theta)] . \end{equation}

The radiation interacts with a QD and produces a force acting upon
the dot. For the case of small QDs compared to the light wavelength
(typical QD sizes are a few nm while the light wavelength in a
colloid is few hundred of nm), the projections of the optical force
acting upon the dot located at the $\textbf{r}$ point are given
by~\cite{Ashkin1975,Chaumet2000}
\begin{eqnarray}\label{force}
F_{j}(\omega)=\frac{1}{2} \sum_{j'} {\rm Re} \left[ \alpha \,
E_{j'}(\textbf{r}) \frac{\partial E_{j'}^*(\textbf{r})}{\partial
r_{j}} \right] ,
\end{eqnarray}
where $j$ and $j'$ run over the Cartesian coordinates $x$, $y$, and
$z$, $E_{j}(\textbf{r})$ is the $j$-projection of the electric field
in the colloid $\textbf{E}(\textbf{r})$ given by Eq.~\eqref{field},
and $\alpha$ is the QD polarizability. The optical force is
proportional to the QD polarizability and scales quadratically with
the electric field. This is because the force originates from the
interaction of the oscillating electric dipole moment of the QD,
induced in turn by the ac electric field and determined by the dot
polarizability, with the ac electric field. In the regime of weak
light-matter coupling, which we consider, for the electric field in
Eq.~\eqref{force} it is sufficient to use the unperturbed field of
the laser beams~\eqref{field}. We note, however, that the optical
force requires more sophisticated self-consistent calculations if
the particle itself considerably affects the electric field
distribution as it can happen, e.g., in the case of strong light
confinement~\cite{Juan2009,Rubin2011,Rubin2011_1,Sadreev2016}.

The dielectric response of the QD caused by exciton transitions within the dot
has a resonant behavior and the polarizability has the form~\cite{Ivchenko_book}
\begin{equation}\label{alpha}
\alpha(\omega,q)=\frac{\pi
a_{B}^{3}\omega_{LT}}{\omega_{0}-\omega-i(\Gamma + q^{3}a_{B}^{3}\omega_{LT}/6)} ,
\end{equation}
where $\omega_0$ is the resonance frequency corresponding to the
exciton transitions, $a_{B}$ is the Bohr radius of the exciton,
$\omega_{LT}$ is the frequency corresponding to the
longitudinal-transverse splitting of exciton states in the host
semiconductor of the QD, which determines the strength of
light-matter coupling in semiconductors at inter-band optical
transitions. The radiative decay rate of excitons is described by
the term $q^{3}a_{B}^{3}\omega_{LT}/6$, the parameter $\Gamma$
stands for the non-radiative decay rate of excitons. The background
dielectric contrast of the QD material and the liquid substance is
neglected.

Straightforward calculations show that the optical force has two
contributions: $\textbf{F} = \textbf{F}^{({\rm gr})} +
\textbf{F}^{({\rm sc})}$. The first (gradient) contribution
$\textbf{F}^{({\rm gr})}$ is determined by the real part of the QD
polarizability. Substitution of Eq.~\eqref{field} for the electric
field in Eq.~\eqref{force} yields
\begin{eqnarray}
F_{x}^{({\rm gr})} = - q \sin \theta \, ({\rm Re} \alpha) E_{1}
E_{2} \sin( 2 x q \sin \theta).
\end{eqnarray}
Taking into account Eq.~\eqref{alpha} for the QD polarizability and
Eq.~\eqref{intensity} for the intensity of radiation, we obtain
\begin{eqnarray}
F_{x}^{({\rm gr})} = \frac{2 \pi^2 }{c n_{\omega}} \frac{a_B^3 \,
\omega_{LT} (\omega_0 - \omega)}{(\omega_0 - \omega)^2 + (\Gamma +
q^{3}a_{B}^{3}\omega_{LT}/6)^2} \frac{d I}{d x} . \label{F_grad}
\end{eqnarray}
This contribution to the optical force is directed along or opposite
to the light intensity gradient depending on the detuning between
the laser field frequency and the exciton frequency.

The second contribution $\textbf{F}^{({\rm sc})}$, commonly referred
to as the scattering and absorbing force, is determined by the
imaginary part of the QD polarizability and has non-zero both
projections. They can be obtained by the substitution of
Eqs.~\eqref{field} and~\eqref{alpha} in Eq.~\eqref{force}, which
yields
\begin{eqnarray}\label{F_scat_x}
F_{x}^{({\rm sc})}  &=&  \frac{q \sin \theta}{2} ({\rm Im} \alpha) [E_1^2 - E_2^2]  \\
&=& \frac{q \sin \theta}{2} \frac{\pi a_B^3 \, \omega_{LT} (\Gamma +
q^{3}a_{B}^{3}\omega_{LT}/6)}{(\omega_0 - \omega)^2 + (\Gamma +
q^{3}a_{B}^{3}\omega_{LT}/6)^2} [E_1^2 - E_2^2] \,, \nonumber
\end{eqnarray}
\begin{eqnarray}\label{F_scat_z}
 F_{z}^{({\rm sc})}  &=&  \frac{q \cos \theta}{2} ({\rm Im} \alpha)
[E_1^2 + E_2^2 + 2 E_1 E_2 \cos(2 x q \sin \theta)] \nonumber \\
 &=& \frac{4 \pi^2 q \cos \theta }{c n_{\omega}}
\frac{a_B^3 \, \omega_{LT} (\Gamma +
q^{3}a_{B}^{3}\omega_{LT}/6)}{(\omega_0 - \omega)^2 + (\Gamma +
q^{3}a_{B}^{3}\omega_{LT}/6)^2} I \,.
\end{eqnarray}
The scattering contribution to the total optical force reaches a maximum at the frequency of the exciton resonance.

We consider in what follows that the amplitudes of the incident laser beams are equal to each other, i.e., $E_1 = E_2$. In this case, $F_{x}^{({\rm sc})}$ vanishes and the scattering force points along the line of the equal intensity of radiation. The total force can be presented in the form
\begin{equation}\label{F_total}
\textbf{F} = F_{0 x} \sin(k x) \hat{\textbf{x}} + F_{0z} [1 + \cos (k x)] \hat{\textbf{z}} \,,
\end{equation}
where
\begin{eqnarray}
F_{0 x} &=& - \frac{4 \pi^2}{c n_{\omega}} \frac{a_B^3 \, \omega_{LT} (\omega_0 - \omega) \, q \sin\theta}{(\omega_0 - \omega)^2 + (\Gamma + q^{3}a_{B}^{3}\omega_{LT}/6)^2} I_0 \nonumber \,, \\
F_{0 z} &=& \frac{4 \pi^2}{c n_{\omega}} \frac{a_B^3 \, \omega_{LT} (\Gamma + q^{3}a_{B}^{3}\omega_{LT}/6) \, q \cos\theta }{(\omega_0 - \omega)^2 + (\Gamma + q^{3}a_{B}^{3}\omega_{LT}/6)^2} I_0 \,, \label{F_ampl}
\end{eqnarray}
$k = 2q \sin\theta$ and $I_0 = (c n_{\omega} / 4 \pi) E_1^2$. For
the parameters $\hbar \omega_0 = 2.35$\,eV, $a_B^3 \omega_{LT} =
0.25 \cdot 10^{-6}$\,cm$^3$/s (Ref.\onlinecite{Ivchenko_book}),
$\hbar \Gamma = 30$\,$\mu$eV (Ref. \onlinecite{Fernee2014}) and $q^3
a_B^3 \omega_{LT} /6 \ll \Gamma$ corresponding to CdSe quantum dots,
the optical force $F_0$ can be estimated as $0.1$\,fN for the
radiation intensity $I_0 = 1$\,kW/cm$^2$.

\subsection{Circulating currents and separation of dots}

In the presence of the scattering contribution, the total optical
force is non-conservative (does not conserve mechanical energy) and
cannot be described as a gradient of the light-induced potential.
Mathematically, it follows from the fact that $\nabla \times \,
\textbf{F} \neq 0$ for the force $\textbf{F}$ given by
Eq.~\eqref{F_total} and the net work done by the optical force in
moving a dot around a closed loop is non-zero. Therefore, the
optical grating of a colloidal solutions leads not only to the
formation of spatially periodic density of the dots but also to the
emergence of spatially periodic circulating fluxes of the dots. The
emergence of a Brownian vortex of a similar nature in a single-beam
optical tweezer has been demonstrated in Ref.~\onlinecite{Sun2009}.
The one-dimensional motion of dielectric particles in an viscous
medium in an optical waveguide was considered in
Ref.~\onlinecite{Sadreev2016}.

The kinetics of quantum dots in the solution is governed by the interplay of the optical forces and the forces of
viscous friction as well as the processes of dot diffusion. The local concentration of the dots $n(x,z,t)$ and the dot flux
$\textbf{j}(x,z,t)$ are related by the continuity equation
\begin{eqnarray}\label{cont_eq}
\frac{\partial n}{\partial t} + \nabla \cdot \, \textbf{j} = 0 \,.
\end{eqnarray}
The dots flux consists of the drift and diffusion terms,
\begin{eqnarray}\label{j_tot}
\textbf{j} = \textbf{j}^{({\rm drift})} + \textbf{j}^{({\rm diff})} \,.
\end{eqnarray}
The drift term is proportional to the optical force driving the dots and
given by
\begin{eqnarray}\label{j_drift}
\textbf{j}^{({\rm drift})} = \mu n \textbf{F} \,,
\end{eqnarray}
where $\mu$ is the mobility of dots in the solution.
The diffusion term is determined by the spatial inhomogeneity of the dot concentration and
has the form
\begin{eqnarray}\label{j_diff}
\textbf{j}^{({\rm diff})} = - D \, \nabla \,n \,,
\end{eqnarray}
where $D$ is the diffusion coefficient. In thermal equilibrium, the mobility and the diffusion coefficient
are connected with each other by the Einstein relation $D = \mu k_B T$, where $k_B$ is the Boltzmann constant
and $T$ is the temperature. The diffusion coefficient of quantum dots with radii $15$ - $20$\,nm in the glycerol/water
solutions with the viscosity $45$ - $55$\,cP  at room temperature was experimentally determined to be $0.5$ - $0.7$\,$\mu$m$^{2}$/s, Ref.~\onlinecite{Lessard2007}. These experimental values were found to be in a good agreement with the Stokes-Einstein relation for the diffusion coefficient of spherical particles in a liquid with a low Reynolds number.

To solve the equation set~\eqref{cont_eq}-\eqref{j_diff} we present the local concentration of the dots
as the sum $n = n_0 + \delta n$, with $n_0$ being the average concentration and $\delta n$ being the correction.
Then, in the regime linear in the light intensity, Eqs.~\eqref{cont_eq}-\eqref{j_diff} yield
\begin{equation}\label{delta_n}
\frac{\partial \, \delta n}{\partial t} - D \Delta \delta n = - \mu
n_0 \, \nabla\cdot \textbf{F}
\end{equation}
and
\begin{eqnarray}\label{j_tot2}
\textbf{j} = \mu n_0 \, \textbf{F} - D \, \nabla  \,\delta n \:.
\end{eqnarray}

Equations~\eqref{delta_n} and~\eqref{j_tot2} are to be solved with boundary conditions. We consider the boundary conditions of
zero fluxes through the cell bottom and top, i.e.,
\begin{equation}
j_z |_{z=\pm h/2} = 0 \,,
\end{equation}
where $h$ is the cell depth.

For the particular form of the optical force given by Eq.~\eqref{F_total}, the solution of the diffusion Eq.~\eqref{delta_n} can be presented in the form
\begin{equation}\label{dn_exp}
\delta n(x,z,t) = a(z,t) + b(z,t) \cos (k x) \,.
\end{equation}
This decomposition leads to the equations
\begin{equation}\label{ab}
\frac{\partial a}{\partial t} - D \frac{d^2 a}{dz^2} = 0 , \;\;  \frac{\partial b}{\partial t} - D \left( \frac{d^2 b}{dz^2} - k^2 b \right) = - \mu n_0 F_{0x} k \,,
\end{equation}
and the boundary conditions
\begin{equation}\label{ab_boundary}
\left. \frac{d a}{dz} \right|_{z=\pm h/2} = \frac{n_0 F_{0z}}{k_B T}  , \;\; \left. \frac{d b}{dz} \right|_{z=\pm h/2} = \frac{n_0 F_{0z}}{k_B T} \,. \;\;\;
\end{equation}
where we took into account the relation $D = \mu k_B T$.

Equations~\eqref{ab} with the boundary conditions~\eqref{ab_boundary} and an arbitrary initial distribution of quantum dots in the solution can be solved by the Laplace transform method. We assume that the solution was initially in thermal equilibrium and then, at $t=0$, the laser radiation is switched on. In this case, the functions $a(z,t)$ and $b(z,t)$ at $t \geq 0$ are given
\begin{eqnarray}\label{a_and_b}
a(z,t) &=& \frac{n_0 F_{0z}}{k_B T} \left[ z
- \frac{4}{h} \sum_{n=0}^{\infty} \frac{(-1)^n \sin k_n z}{k_n^2} {\rm e}^{-D k_n^2 t} \right] ,  \\
b(z,t) &=& - \frac{n_0 F_{0x}}{k_B T \, k} \left[ 1 - {\rm e}^{-D k^2 t} \right] + \frac{n_0 F_{0z}}{k_B T \, k} \nonumber \\
&\times& \left[ \frac{\sinh kz}{\cosh (kh/2)}
- \frac{4}{h} \sum_{n=0}^{\infty} \frac{(-1)^n \sin k_n z}{k^2 + k_n^2} {\rm e}^{-D (k^2 +k_n^2) t} \right] , \nonumber
\end{eqnarray}
where $k_n = (\pi / h)(1+2n)$. The maps of the dot concentration $\delta n(x,z,t)$ and the dot fluxes $\textbf{j}(x,z,t)$ can be readily found from Eq.~\eqref{dn_exp} and~\eqref{j_tot2}, respectively.

The steady-state distributions of the dot concentration and the dot fluxes induced by cw radiation can be obtained by considering the limit
of the functions $a(z,t)$ and $b(z,t)$ at $t \rightarrow \infty$. This yields
\begin{equation}\label{delta_n_stat}
\delta n =  \frac{n_0 F_{0z}}{k_B T} z + \frac{n_0}{k_B T} \left[ F_{0z} \frac{\sinh (k z)}{\cosh (k h /2)} - F_{0x} \right]
\frac{\cos k x}{k}
\end{equation}
and
\begin{eqnarray}\label{j_stat}
j_x &=& \mu n_0 F_{0z} \frac{\sinh (k z)}{\cosh (k h /2)} \sin (k x) \,, \\
j_z &=& \mu n_0 F_{0z} \left[1 - \frac{\cosh(kz)}{\cosh (k h /2)} \right] \cos (k x) \,. \nonumber
\end{eqnarray}
The steady-state spatial modulation of dot concentration is produced by both $F_x$ and $F_z$ components of the optical force while
the steady-state fluxes are induced by the non-conservative $F_z$ component only.

\section{Results and discussion}

\begin{figure}
\includegraphics[width=85mm]{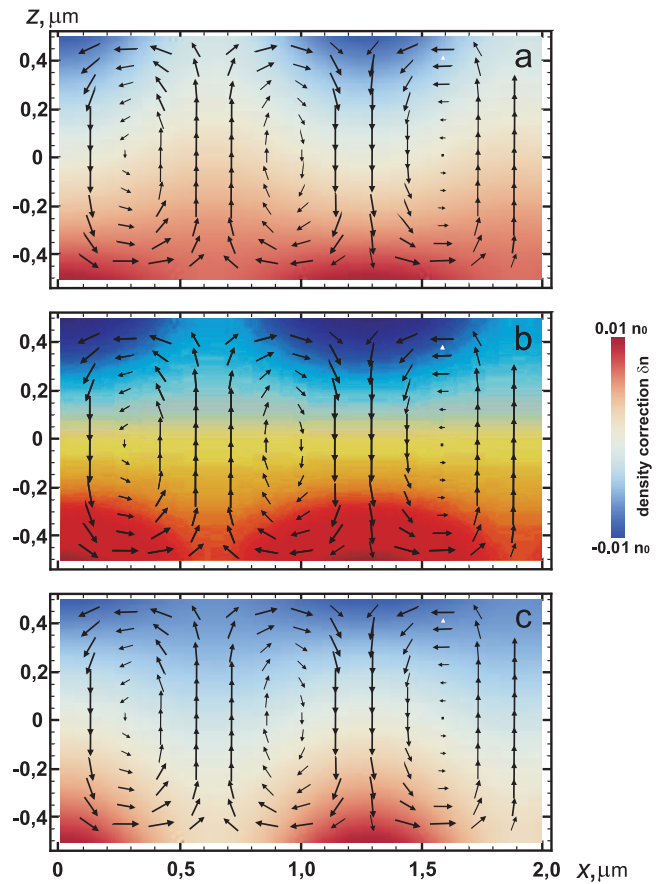}%
\caption{(Color online) Steady-state distributions of the dot density (color map) and the dot circulating
fluxes (arrows) in a colloidal solution induced by optical grating, as sketched in Fig.\ref{figure1}.
Panels a, b, and c correspond to the negative detuning $\omega_0-\omega=-3\Gamma$, zero detuning $\omega_0=\omega$,
and positive detuning $\omega_0-\omega=3\Gamma$, respectively. All the distributions are calculated after Eqs.~\eqref{delta_n_stat}
and~\eqref{j_stat} for the parameters $\hbar \omega_0=2.35$\,eV, $\hbar \Gamma = 30$\,$\mu$eV (Ref.~\onlinecite{Fernee2014}), $a_B^3 \omega_{LT} = 0.25 \cdot 10^{-6}$\,cm$^3$/s (Ref.~\onlinecite{Ivchenko_book}) corresponding to CdSe quantum dots, $T=300$\,K, $k = 0.5 \cdot 10^5$\,cm$^{-1}$, $\theta=8^\circ$, $h=1$\,$\mu$m, and $I_0 = 1$\,kW/cm$^2$.}
\label{figure2}
\end{figure}

Figure~\ref{figure2} demonstrates the steady-state maps of the
quantum dot density $n(x,z)$ and the dot circulating fluxes
$\textbf{j}(x,z)$ induced by the optical grating of a colloidal
solution. The density distribution $n(x,z)$ is encoded by color, the
flux distribution $\textbf{j}(x,z)$ is shown by black arrows. Panels
a, b, and c correspond to a negative, zero, and a positive detuning
$\omega_0 - \omega$ between the dot resonant frequency $\omega_0$
and the radiation frequency $\omega$. Typically, the resonant
frequencies of the dots in a solution have a dispersion due to,
e.g., the dispersion of the dot sizes. The optical grating leads to
the formation of a spatially periodic density of the dots in the
longitudinal direction $x$ and inhomogeneous distribution of the
dots in the vertical direction $z$. The modulation $\delta n /n_0$
is about $1 \%$ for the radiation intensity $1$\,kW/cm$^{2}$. At
fixed temperature and radiation intensity, the modulation of the QD
density is determined by the frequency detuning $\omega_0 - \omega$,
the radiative and non-radiative decay rates of excitons, $\Gamma$
and $q^3a_{B}^3 \omega_{LT}/6$, respectively. The non-radiative
decay rate is typically much larger than the radiative decay rate at
room temperature. Therefore, the higher degree of the QD density
modulation and the higher spectral sensitivity of the QDs are
expected for QDs with suppressed non-radiative channels of exciton
recombination.

Quantum dots with the exciton frequencies detuned from the radiation frequency
(panels a and c) are acted upon by both the scattering component $F_z$ and the gradient component $F_x$ of the optical force.
The scattering component is directed along the $z$ axis but its magnitude depends on the
longitudinal coordinate $x$, see Eq.~\eqref{F_ampl}. It leads to the formation of a vertical profile of the dot density
and a spatially periodic dot density in the longitudinal direction at the cell top and bottom. More importantly, the scattering force
gives rise to persistent circulating fluxes of the dots shown by black arrows. The circulating fluxes are more pronounced in cells with the depth $h$ comparable to and exceeding the wavelength $2\pi /k$ since in shallower cells the vertical motion of the dots in suppressed. The gradient component $F_x$ leads to the formation of a spatially periodic density of the dots in the longitudinal direction.
This effect is most clearly visible at $z=0$. The direction of the gradient force depends on the sign of the detuning, therefore the density modulation at $z=0$ is of opposite signs for the dots with positive (panel a) and negative (panel c) detuning.

For quantum dots with the resonant frequency corresponding to the radiation frequency (panel b), the gradient component $F_x$
vanishes and the kinetics is solely determined by the scattering force $F_z$. The scattering force gives rise to the inhomogeneous
distribution of the dots as well as the circulating fluxes of the dots.

\begin{figure}
\includegraphics[width=85mm]{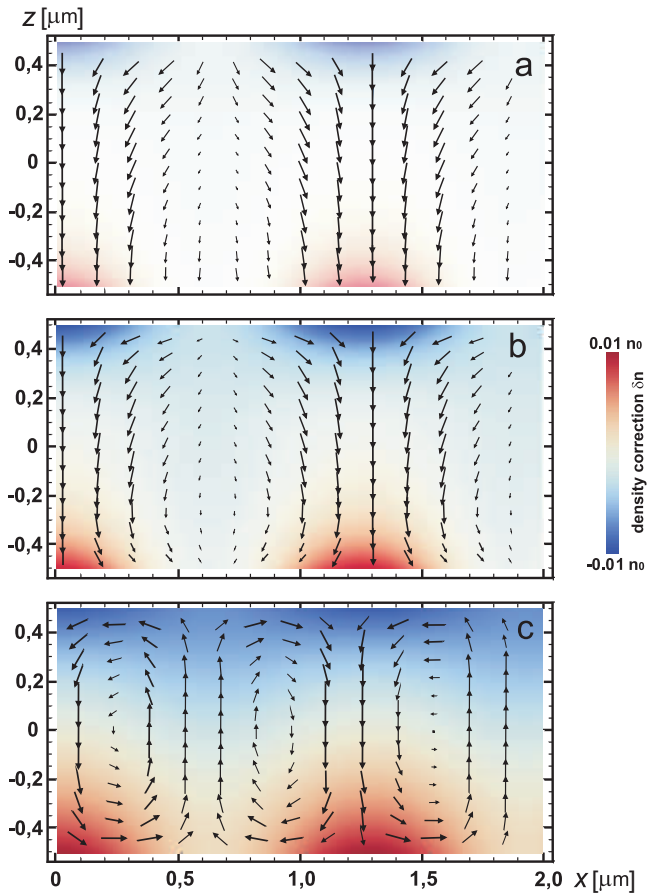}%
\caption{(Color online) Snapshots of the distributions of the dot
density (color map) and the dot fluxes (arrows) at the times (a)
$t=0.002$\,s, (b) $t=0.02$\,s, and (c) $t=0.1$\,s after the
radiation is switched on. The distribution are calculated for the
parameters of Fig.~\ref{figure2}c and the diffusion coefficient $D =
0.5 \cdot 10^{-8}$\,cm$^{2}$/s.} \label{figure3}
\end{figure}

Figure~\ref{figure3} shows some stages of the steady-state
distribution formation after the laser radiation is switched on. As
it follows from Eqs.~\eqref{a_and_b}, the time scale of the
formation of longitudinal periodic density of the dots is determined
by the time $\tau = 1/(D k^2)$. For the wave vector $k= 0.5 \cdot
10^{4}$\,cm$^{-2}$ and the diffusion coefficient $D = 0.5 \cdot
10^{-8}$\,cm$^{2}$/s, this time $\tau \approx 0.1$\,s.
Figures~\ref{figure3}a,~\ref{figure3}b, and~\ref{figure3}c show the
snapshots of the density and flux distributions at $t=0.002$\,s,
$0.02$\,s, and $0.1$\,s, respectively. Just after the radiation is
switched on, the dots are homogeneously distributed in the solution.
At this stage, the dot fluxes are completely determined by optical
forces since there are no contributions related to the dot density
gradients. At large times, the density gradients are formed, the
fluxes become closed, and the distributions approach the
steady-state ones.

\section{Summary}

We have presented a theoretical study of optical forces acting on
semiconductor quantum dots in a colloid and formation of a fluid
photonic crystal from the dots by optical grating. In such a system,
the density of the dots and the fluxes of the dots are periodically
modulated in space suggesting the periodic modulation of the optical
properties of the colloid. By considering the interplay of the
optical forces and the processes of viscous friction and diffusion
of dots, we have calculated the steady-state distributions of the
dot density and fluxes as well as the time scales of the fluid
photonic crystal formation. The results can be employed for creating
dynamic photonic structures with tunable parameters and further
study of their optical properties. Other possible applications
include optically-induced mixing of colloids and the separation of
quantum dots with different resonant frequencies.

This work was supported by the RF President Grant for young
scientists MD-4550.2016.2 and the RFBR grant 16-02-00204.

 \pagebreak

\end{document}